Intrinsic resistance peaks in AB-stacked multilayer graphene with odd number of layers

Tomoaki Nakasuga[1], Taiki Hirahara[1], Kota Horii[1], Ryoya Ebisuoka[1], Shingo Tajima[1], Kenji Watanabe[2], Takashi Taniguchi[2] and Ryuta Yagi[1]

[1]Graduate School of Advanced Sciences of Matter (AdSM), Hiroshima University, Higashi-Hiroshima 739-8530, Japan,
[2]National Institute for Materials Sciences (NIMS), Tsukuba 305-0044, Japan.

ABSTRACT

The intrinsic resistance peak (ridge) structures were recently found to appear in the carrier density dependence plot of the resistance of the AB-stacked multilayer graphene with even numbers of layers. The ridges are due to topological changes in the Fermi surface. Here, these structures were studied in AB-stacked multilayer graphene with odd numbers of layers (5 and 7 layers) by performing experiments using encapsulated high-quality graphene samples equipped with top and bottom gate electrodes. The intrinsic resistance peaks that appeared on maps plotted with respect to the carrier density and perpendicular electric field showed particular patterns depending on graphene's crystallographic structure, and were qualitatively different from those of graphene with even numbers of layers. Numerical calculations of the dispersion relation and semi-classical resistivity using information based on the Landau level structure determined by the magnetoresistance oscillations, revealed that the difference stemmed from the even-odd layer-number effect on the electronic band structure.

(Corresponding Author     yagi@hiroshima-u.ac.jp)

# 1 Introduction

Since the discovery of massless Dirac fermions in graphene, a number of scientific investigations [1-3] have tried to elucidate its physical property and potential applicability. The electronic properties of graphene depend strongly on the crystallographic structure. While monolayer graphene has a single band with a massless dispersion relation [4-6], bilayer graphene has with a massive dispersion relation [4,7-10]. As the number of layers increases, many different stacking structures become possible, each of which is expected to have a particular band structure. In particular, AB-stacked graphene shows regularity in the evolution of its band structure. $2N$ ($N$: integer) layer graphene has $N$ bilayer-like band(s), and $2N+1$ layer graphene has $N$ bilayer-like band(s) and a monolayer-like band, as shown in Fig. 1a [8,11-17]. Detailed band calculations have suggested that the band structure of multilayer graphene is much more complicated than those shown in Fig. 1 a. The band structure of graphene in the high-energy regime has been studied in optical spectroscopic experiments [18-24]. Moreover, while the low-energy band structure, which affects transport phenomena, has not been fully revealed by optical measurements, the advent of techniques for making high-quality graphene samples has made it possible to probe the low-energy band structure by using Shubnikov- de Haas oscillations [4,25-35].

Recently, high-quality multilayer graphene was found to show intrinsic resistance peaks in its carrier density dependence [31,33,36]. Detailed measurements using graphene samples with top and bottom gate electrodes have uncovered intrinsic resistance ridges structure specific to the band structure of AB-stacked 4-layer [31,33] and 6-layer graphene [36]. These intrinsic resistance ridges are considered to be a promising means of probing the band structure of two-dimensional materials. The ridges (or peaks) are related to the topological changes in the Fermi surface [31]; the complicated dispersion relations of multilayer graphene are tunable with a perpendicular electric field, and the resultant shape of the Fermi surface (energy contour of the dispersion relations) varies depending on the chemical potential. The ridges in AB-stacked graphene with even number of layers graphene are principally due to two reasons [33,36]. One is the nearly flat band structure created at the bottoms of the bilayer-like band by the perpendicular electric field; the field opens a band gap at the bottoms of these bands, and make them approximately flat. This results in a the heavy band mass [7,10,37-40]. Conspicuous split ridges appear on the map of resistivity as a function of carrier density and perpendicular electric flux density. The other reason is formation of mini-Dirac cones [41], which are created by the perpendicular electric field [31,33,36]. Trigonal warping locally closes the energy gap created by the perpendicular electric field, and thereby, mini-Dirac cones are created. Mini-Dirac points appear as sharp resistance ridges. In AB-stacked 4-layer graphene, they appear at the charge neutrality point [31,33], while and in AB-stacked 6-layer graphene, they appear not only at the charge neutrality point but also at non-zero carrier densities [36]. In this paper we will examine the intrinsic resistance peak structure of AB-stacked multilayer graphene with odd numbers of layers. We found that the resistance ridges are qualitatively different from graphene with even number of layers. We will show that this difference originates from the dispersion of the bilayer-like band which is hybridized with the monolayer-like band. The bottoms of the bilayer-like bands form particular structure that is qualitatively different from that of the even numbers of layers.

## 2 Experimental

Our samples consisted of high-quality graphene flakes encapsulated with thin flakes of *h*-BN, and equipped with top and bottom gate electrodes, as shown schematically in Fig. 1 a and b. The graphene flakes were prepared by mechanical exfoliation of high-quality Kish graphite crystals. Thin *h*-BN flakes were also prepared using a similar method. A stack consisting of graphene layers and *h*-BN layers was formed by using the transfer technique described in Ref. [42] and Ref [25]. The graphene sample was formed on a Si substrate (covered with $SiO_2$) which was heavily doped and remained conducting at low temperature. The substrate served as the bottom gate electrode. The top gate was formed by transferring graphene with a few layers onto the top of the encapsulated graphene. The resulting stack, consisting of the graphene and *h*-BN flakes, was patterned into a Hall bar by using reactive ion etching with a low pressure mixture of $CF_4$ and $O_2$ gas. Therefore, the top gate electrode and the effective sample area had an exact geometry (Fig. 1(b) and 1(c)). Electrical contact with graphene was attained by using the edge contact technique reported in Ref. [42]. Electrical contact with top gate electrode was carefully made at a point where the graphene sample to be measured was not underneath the *h*-BN, so as not to make a direct connection with the graphene to be measured.

The number of layers and the stacking of the graphene were verified by various methods. We identified their effect on the characteristic Landau level structures [17], which are specific to a particular number of layers and stackings (see the Appendix).

3. Results and Discussion

(1) AB-stacked 5-layer graphene

AB-stacked 5-layer graphene is a typical example of odd layer multilayer graphene with the multiple bilayer like bands. It is expected to have two bilayer-like bands and a monolayer band [11-13,16-17,43-45]. Figure 1(d) shows the back gate voltage ($V_b$) dependence of the resistivity for different top gate voltages ($V_t$), which was measured at $T = 4.2$ K. The mobility ($\mu$) was calculated with the simple formula $\mu = 1/|n_{tot}e\rho|$ using data for the $V_b$ dependence of resistivity ($\rho$) with $V_t = 0$. It was about $1.9 \times 10^5$ cm²/Vs in the electron regime, and $1.1 \times 10^5$ cm²/Vs in the hole regime at large carrier densities. Here, $n_{tot}$ is the total carrier density carrier density. It is clear that varying the top gate changed the overall shape of the resistance traces with respect to $V_b$. The data with $V_t = 0$ shows conspicuous double-peak structures whose peak resistivities are approximately the same. These structures would originate from the bottoms of the bilayer-like bands [31,33]. With increasing $|V_{tg}|$, the shapes of the traces change into ones with a main peak and a small side peak. The resistivity of the main peak increases with $|V_{tg}|$, until it saturates and then slightly decreases for large $|V_{tg}|$. This behavior is reminiscent that of AB-stacked 4-layer [31,33] and 6-layer graphene [36] and is strikingly different from the behavior of bilayer [46-49] or AB-stacked trilayer graphene[49-53]. Bilayer graphene shows insulating behavior as the top gate voltage increases, while trilayer graphene shows the opposite behavior; the resistivity of the peaks appearing near the charge neutrality point decreases with increasing top gate voltage.

To investigate the above mentioned properties further, we measured the top and bottom gate voltage dependence of the resistivity in detail. The results are summarized in the map of resistivity with respect to $V_t$ and $V_b$, as shown in Fig 2(a). Resistivity peaks appear as ridges. Salient resistance ridges on a linear line from the upper left to lower right satisfy the condition of charge neutrality, which corresponds to the large peaks in Fig 2(d) for $|V_t| > 0$. In addition, side peaks which are parabolic in shape are discernible in the figure. Because the peak structure would result from a variation in the dispersion relation arising from the perpendicular electric field as in AB-stacked 4- and 6-layer graphene [31,33,36], we replotted the map as a function of total carrier density ($n_{tot}$) and electric flux density ($D_\perp$) perpendicular to the graphene. The total carrier density can be

calculated by summing the carrier densities induced by the top and bottom gate voltages as

$$n_{tot} = (C_t(V_t - V_{t0}) + C_b(V_b - V_{b0}))/(e). \quad (1)$$

Here, $C_t$ and $C_b$ are the specific capacitances of the top and bottom gate electrodes, respectively. $V_{t0}$ and $V_{b0}$ represent the shift in gate voltage due to carrier doping associated with the top and bottom gate electrodes. The effect of the perpendicular electric field can be estimated using electric flux density induced by the top and bottom gate voltages, which is given by

$$D_\perp = (C_t(V_t - V_{t0}) - C_b(V_b - V_{b0})) / 2. \quad (2)$$

From the charge neutrality condition in Fig. 2(a), one can estimate the ratio of the capacitances ($= V_t/V_b$) to be about 3.8. The specific capacitances were calculated from the Landau level structure measured under the condition, $D_\perp=0$, to be $C_t = 395$ $aF/\mu m^2$ and $C_b = 104$ $aF/\mu m^2$. Figures 2(b) and 2(c) show maps of $\rho$ and $d\rho/dn_{tot}$ as a function of $n_{tot}$ and $D_\perp$. In these figures, parabolic ridges are discernible for both the electron and hole regimes. Similar but more complicated parabolic ridge structures were also observed in AB-stacked 4 [31,33] and 6-layer graphene [34,36].

The dispersion relations in the absence and presence of perpendicular electric fields were numerically calculated to examine the relation between the band structure and the resistance ridge structure in the 5-layer graphene. The calculation was based on the effective mass approximation and the Slonczewski-Weiss-McClure (SWMcC) parameters [54-56] of graphite were used. Screening of induced carriers was taken into account by using the distribution of induced carriers in graphene. We assumed that carriers carrier in each layer decay exponentially with a decay length of λ, which is roughly consistent with the results of the Thomas-Fermi approximation [57]. In the calculation we took λ = 0.45 nm, approximately the same value expected from a self-consistent calculation of the screening length [58], and approximately the same as the experimental value obtained from Landau-level structures in multilayer graphene [17]. Figure 3 (a) shows the dispersion relations of the AB-stacked 5-layer graphene numerically calculated for different values of $|D_\perp|$. The dispersion relations for AB-stacked 4 layer graphene are shown in Fig. 3(b) for comparison. For the 5-layer graphene, there are two sets of bilayer-like band and a monolayer-like band, which are complicatedly hybridized near $E = 0$ [35] (more complicated than what is shown in Fig. 1(a)). Applying a perpendicular electric field opens energy gaps., i.e., differences in energy between the bottoms of the bands. The gaps increase with increasing $|D_\perp|$; thereby, the dispersion relations look rather simplified. For convenience, we labeled the band as $\alpha_e$—$\gamma_h$, and the bottoms of

the bands a, b, c, b', and a'. Bands $\gamma_e$ and $\gamma_h$ in the 5-layer graphene are monolayer-like bands. The remaining bands are principally bilayer-like bands, as in the AB-stacked 4-layer graphene, but they differ significantly between the 4- and 5- layer cases. In particular, the energy gap between $\alpha_e$ and $\beta_e$ and the one between $\alpha_h$ and $\beta_h$ are significantly larger in the 4-layer graphene than in the 5-layer graphene. It can be seen that the structures of band $\alpha_e$ and $\alpha_h$ in the 5-layer graphene are more complicated than those in the 4-layer graphene; this difference would originate from the hybridization with the monolayer-like band in the 5-layer graphene.

The difference in the band structure results in particular resistance ridge structures. The characteristic band positions are closely related to the resistance ridges. We have calculated the semi-classical resistivity based on the Boltzmann equation with the constant relaxation-time approximation. (The calculation is similar to the one performed on AB-stacked 4-layer graphene [31]). We took into account possible energy broadening due to scattering. Figure 4(a) compares the experimental and calculated maps of $d\rho/dn_{tot}$ plotted as a function of $n_{tot}$ and $D_\perp$. It can be seen that the calculations approximately reproduced the experimental result. Conspicuous ridge structures are labeled with the characteristic positions of the band structure in Fig 3(a). Ridge c stems from the mini-Dirac cones formed at the charge neutrality point. Ridges b and and b' are for the bottoms of the bilayer-like bands $\beta_e$ and $\beta_h$. As for positions a and a', which correspond to the bottoms of the monolayer-like bands, structures hardly appeared in the experimental results, possibly because the variation in the conductivity was rather smaller than at the other characteristic positions in the bands. As shown in Fig. 4(b), the structures are barely visible in the simulation with reduced energy broadening.

Now we let us discuss the differences between the 5-layer and 4-layer cases. The resistance ridges of the 5-layer graphene, which are parabolic in shape, are qualitatively different from those of the 4-layer graphene. The ridges in the 4-layer case show clear splitting with increasing $|D_\perp|$ [31,33]. This is due to formation of an energy gap between the bilayer like bands, as shown in Fig. 3(b). On the other hand, in the 5-layer graphene, the bottom of $\beta_e$ almost touches $\alpha_e$, and the bottom of $\beta_h$ has approximately the same energy as the local bottom of $\alpha_h$. This qualitatively different band structure results in the 5-layer graphene not having any split ridge structures for the bilayer-like bands.

Although the 4-layer and 5-layer graphene have significantly different electronic band structures, they showed have similar resistance ridges that appear at $n_{tot} = 0$ for $|D_\perp|$ above $\sim 0.5 \times 10^{-7} \text{cm}^{-2}$As. In both cases, this is because the ridge originates from the formation of mini-Dirac cones [31,36,41] near $E = 0$ for large $|D_\perp|$, as can be seen in Fig. 3(a) and 3(b). In the 5-layer case, three sets of mini-Dirac cones are created at different wave numbers in $k$-space. Among them, the two located at $k_x \neq 0$ (see Figure 3(a)) arise from the bilayer-like band because of trigonal warping. They are both three-fold degenerate in a valley (K or K'). The other set of mini-Dirac cones, which are located at $k_x = 0$, apparently originate from the monolayer-like band. The mini-Dirac cone structure in the 4-layer case is strikingly different. There are large mini-Dirac cones (in positive $k_x$) and a small mini-Dirac cone-like structures (in negative $k_x$) which have gaps. The cones and the small cone-like structure are both three-fold degenerate in the K and K' valley. In the both the 4- and 5-layer cases, perpendicular electric field resulted complicated massive bands changing into linear bands near the charge neutrality point, and thereby, the resistance ridges near the $n_{tot} = 0$ appeared.

The simulation with reduced energy broadening reveals the resistance ridges associated with the monolayer band for the bottoms of the monolayer bands $\gamma_e$ and $\gamma_h$ (Fig. 4(b)). However, they are hardly visible in the experimental data. In sufficiently large perpendicular electric fields, energy gaps are created for the monolayer-like band because of hybridization with bilayer-like bands (Fig. 3(a)). Although the monolayer-like band has a non-zero band mass near the bottoms of the bands, the mass is much smaller than those for the bottoms of the bilayer-like bands. This would make it hard the resistance ridges for the bottoms of $\gamma_e$ and $\gamma_h$.

(2) AB-stacked 7-layer graphene

AB-stacked 7-layer graphene, which has three sets of bilayer-like bands and a monolayer-like band, also shows characteristic resistance ridges for odd numbers of layers. We studied the intrinsic resistance peaks of the 7-layer sample that had a similar structure to that of the 5-layer sample. The mobility at a large carrier density was $\mu = 6.9 \times 10^4$ cm$^2$/Vs in the electron regime and $5.0 \times 10^4$ cm$^2$/Vs in the hole regime. Figure 5(a) shows a map of resistivity as a function of $V_b$ and $V_t$, which was measured at $T = 4.2$ K. The ratio of the specific capacitance was estimated to be $C_t/C_b = 3.98$. $C_t = 446$ aF/µm$^2$ and $C_b = 112$ aF/µm$^2$. Figure 5 (b) is a replot as a function of $n_{tot}$ and $D_\perp$. The resistance ridges are distinct from those of the 4-layer [31,34], 5-

layer and the 6-layer cases [33,36].

Although the 6-and the 7-layer graphene have more complicated band structures than those of 4- and 5-layer graphene, they show characteristic differences in the band structure reflecting the even-odd layer number effect. Figure 6(a) shows the numerically calculated dispersion relation of the 7-layer graphene for some values of $|D_\perp|$, while Fig. 6 (b) shows those for the 6-layer case for comparison. The SWMcC parameters of graphite, and $\lambda = 0.45$ nm were used in the calculation. Bands are labeled $\alpha_e$, $\beta_e$, $\gamma_e$, $\delta_e$, $\alpha_h$, $\beta_h$, $\gamma_h$, and $\delta_h$. Characteristic points in the band diagram are labeled a-d and a'-c'. The dispersions for the 7-layer graphene under a perpendicular electric field are much more complicated than those of the 6-layer graphene because of hybridization of the bilayer-like bands with a monolayer-like band, as was seen earlier for the cases of the 4- and 5-layer graphene. In the 6-layer graphene, the application of a perpendicular electric field opens energy gaps between the bilayer-like bands, and the dispersion relations are nearly flat near the bottoms of each band. On the other hand, no such flat dispersion relations form in the 7-layer graphene. The bottoms $\gamma_e$ and $\beta_e$ ($\gamma_h$ and $\beta_h$) nearly make contact with the small energy gaps. The structures apparently originate from hybridization with the monolayer-like band. On the other hand, for large $|E|$, one can see that bands $\delta_e$ and $\delta_h$, which originate from the monolayer-like band, have rather simple shapes.

To see the correspondence of the intrinsic resistance ridges to the dispersion relations, we compared the experimental results with the numerically calculated resistivities (Fig. 7).It is clear that the theoretical results approximately explain the experimental results. Resistance ridges appear at the corresponding positions in the band structures. First, let us examine the ridges appearing in the vicinity of $n_{tot} = 0$. One can recognize the resistance ridge near the charge neutrality condition as in the 4-layer, 5-layer, and 6-layer graphene. Comparing the experimental results with those of the band calculation, it can be seen that mini-Dirac cones are created at points d in the vicinity of the charge neutrality point for large $|D_\perp|$. In the AB-stacked 7-layer graphene, the dispersion relations at $|D_\perp| = 0$ show a semi-metallic band structure; the electron and hole bands overlap near $E = 0$. Applying a perpendicular electric field created mini-Dirac cones, from which conspicuous ridges formed.

Next, we turn to the other resistance ridges. The bottoms of the bilayer bands b and b' (Fig. 6 (a)) appear as resistance ridges in Fig. 7. Apparently, there are no split ridge structures arising from the bottoms of bilayer-like bands, as in the 5-layer case. Unlike the 5-layer case, conspicuous arising from mini-Dirac cones (indicated by c and c' in Fig. 6(a)) are visible as in the 6-layer case [36], at carrier densities different from charge neutrality. In addition, the experimental results do have clear ridge structures for the monolayer-like band a and a', as in the 5-layer case; the lack should again be due to relatively small carrier density and light band mass.

## 4  Discussion

The even-odd layer-number effect in the band structure is an intrinsic feature of AB-stacked multilayer graphene. This feature can be seen in the Landau level structures: the absence or presence of the Landau levels due to the monolayer-like band [11,13,17,25,28,30-32-35,45,59-60] and the absence or presence of valley splitting at zero perpendicular electric field [25,35]. As for the dispersion relation at zero magnetic field, the low-energy band structures are expected to be rather complicated, and information can be extracted from the Landau level structure indirectly through the band parameters with which the dispersion relation can be calculated. As described in the previous section, the intrinsic resistance ridge structure reflects the specific band structure of graphene, and this allows us to probe the band structure directly in the transport experiments. Here, we address the evolution of the resistance ridge structure in AB-stacked multilayer graphene with increasing layer number. The numerically calculated resistance ridge structures for 4 to 7 layers are summarized in Fig. 8. (The calculation for the 4-layer case is reported in Ref. [31].) It is clear that the resistance ridges (peaks) appear at different positions in the diagram: the ridge structures have a specific pattern depending on the number of layers. One can thus determine the number of layers and stacking by using the diagram. Resistance ridges due to bilayer bands show splitting in graphene with even numbers of layers, while the splitting is absent from graphene with odd numbers of layers. In addition, the resistance ridge due to the monolayer-like band in the graphene with the odd numbers of layers are rather small.

On the other hand, the mini-Dirac points form relatively strong peaks compared with the bottoms of the bands. For example, ridge structures at $n_{tot} = 0$ appear regardless of the number of layers. The 6- and 7-layer graphene show relatively strong peaks at the mini-Dirac points (MDP) at non-zero carrier densities.

On the ridges formed at $n_{tot} = 0$, the resistivity tends to increase with increasing $|D_\perp|$,

but it saturates (and slightly decreases in some cases) at large $|D_\perp|$. The early graphene research reported that bi- and trilayer graphene had different responses to a perpendicular electric field: bilayer graphene becomes insulating because the energy gap opens [47-48,61], while trilayer graphene becomes more metallic [50-52,61] (i.e. its resistivity decreases). However, this sort of behavior does not persist in graphene consisting of more layers, as shown in previous work [31,33,36] and this study. The behavior is consistent with the formation of mini-Dirac cones in the vicinity of the charge neutrality point. AB-stacked 5-7 layer graphenes (and possibly the 4 layer graphene) are semi-metallic near the charge neutrality point in the absence of a perpendicular electric field. Electrons and holes are compensated, so that there would be considerably many number of carriers that contribute to the conductance. The formation of mini-Dirac cones tends to decrease the number of carriers. The absence of insulating behavior can be understood from the minimum conductivity of monolayer graphene at the charge neutrality point. Theory predicts a minimum conductivity of about $e^2/\hbar$ at the Dirac point [4,6,62]. Although a Dirac point is difficult to realize in an actual experiment because of inhomogeneity [63-66], may experiments have shown that there is a minimum conductivity, whose value is not universal.

Summary and concluding remarks

Intrinsic resistance ridge structures of AB-stacked 5- and 7- layer graphene, which appear as a function of carrier density and perpendicular electric field, were studied together with the band structure by using an encapsulated graphene device equipped with top and bottom gate electrodes. We found that the intrinsic resistance peaks (ridges) in multilayer graphene with an odd numbers of layer are strikingly different from the graphene with in an even number of layers: only graphene with an even number of layers show split ridges due to the formation of nearly flat bands. This difference results from hybridization of the bilayer-like band with the monolayer-like band in the graphene with odd number of layers. Thus, these results show that the resistance ridges can be used to probe the electronic band structure of two-dimensional materials.

Appendix

A Determination of number of layers and stacking

The number of layers and their stacking were determined by combined use of atomic force microscopy (AFM) and Raman spectroscopy. In particular, the number of layers and stacking were determined after calibrating the relation between the Raman spectral shape and the number of layers of graphene determined by AFM. The spectral shape of the ABA stacking showed a systematic evolution [17,67-70], that was considerably different from that of ABC stacking [67,69-72]. The details are described in Ref. [17]. We also used the Landau level structures which can be deduced from the Shubnikov-de Haas oscillations in the low-temperature magnetoresistance. The Landau level structures reflects the electronic band structure of graphene directly, meaning that it is one of the most reliable methods to determine the number of layers and stacking. A map of magnetoresistance with respect to the carrier density and magnetic field (Landau fan diagram) reveals graphene's detailed low-energy band structure that is specific to the number of layers and stacking. The number of layers and stacking of the measured samples were verified by referring a list of fan diagrams for AB-stacked graphene with known numbers of layers [17].

B Landau level structure in AB-stacked 5-layer graphene

The AB-stacked 5-layer graphene sample showed Shubnikov-de Haas oscillations in the magnetoresistance which was measured at $T = 4.2$ K. Figures 9(a) and 9(b) show maps of the longitudinal resistivity ($\rho_{xx}$) and its derivative with respect to the magnetic field ($d\rho_{xx}/dB$), plotted as a function of magnetic field $B$ and carrier density $n_{tot}$. Here, $n_{tot}$ was varied by controlling the top and bottom gate voltages so as to satisfy the condition, $D_\perp = 0$. The stripes are Landau levels for particular bands with particular Landau indices. The observed Landau level structure near the charge neutrality point is approximately the same as that in the previous report for AB-stacked 5-layer graphene, which was measured from a sample with a single gate electrode [17]. This confirms that our sample was identified AB-stacked 5-layer graphene, because Landau level structure is the fingerprint of the electronic band structure of graphene.

The overall structure of the Landau levels can be approximately explained by a numerical calculation based on the effective mass approximation. Figure 9(c) shows energy eigenvalues calculated for the Slonczewski-Weiss-McClure parameters which are approximately the same as those of graphite, and Figure 9(d) is the calculated density of states. Although refining the SWMcC parameters would give a better fitting to the experiment, energy gaps with $\nu = -2, 14, 18$ are clearly visible in the experimental data. The filling factor for the gaps satisfies the relation $4(N+1/2)$ with integer $N$, as in the mono-layer graphene. In addition, the Landau levels for the monolayer-like band are visible in Fig 8(b) (indicated by the red bars).

The energy gap with $\nu = -2$ characterizes the AB-stacked 5-layer graphene. It occurs near the charge neutrality point above a few Tesla and appears between the zero-mode Landau levels; no Landau level crossings occur for the larger magnetic field. Similar characteristic energy gap structures appears in AB-stacked multilayer graphene with more layers, and one can identify the number of layers by using the filling factor of the gap. The gap occurs at $\nu = 0$ in the case of AB-stacked graphene [31,33,35], while it appears at $\nu = 4$ in AB-stacked 6 layer [36]. To be shown later, in the 7 layer, it appears at $\nu = 6$.

## C Landau level structure in AB-stacked 7-layer graphene

Figures 10(a) and 10(b) show maps of $R_{xx}$ and $dR_{xx}/dB$ as a function of $n_{tot}$ and $B$. Highly complicated beatings of the Shubnikov-de Haas oscillations can be seen. The energy gaps and the Landau level crossing near the charge neutrality point approximately reproduce the fan diagram measured for single-gated graphene samples [17], which confirms that our sample is the AB-stacked 7-layer graphene. Conspicuous energy gaps appear at $\nu = -6$. Figure 10(c) shows the numerically calculated Landau level spectra for the SWMcC parameters of graphite, while Figure 10(d) is a map of the corresponding density of states. The calculation approximately accounts for the overall Landau level structures and positions of the conspicuous energy gaps. In particular, the energy gap at $\nu = -6$ is visible between the zero-mode Landau levels of the bilayer-like band.

## D Calculation of dispersion relation and Landau levels

The dispersion relations at zero magnetic field were calculated using the Hamiltonian for the effective mass approximation which is based on the tight-binding model [5,35,43,45]. Landau levels were numerically calculated by expanding the wave functions with Landau functions [14,35,59,74-75] and evaluating the eigenvalues of the Hamiltonian. The density of states was calculated by assuming that each Landau level had a carrier density of degeneracy multiple $eB/h$ [35].

The electrostatic potential due to the perpendicular electric field was calculated by taking the screening of each layer into account. Multilayer graphene is atomically thin, as are other two-dimensional materials, so that an externally applied perpendicular electric field is expected to penetrate the graphene but to be shielded layer by layer [17,23,57-58,76-81]. The internal electric field significantly changes the electrostatic potential for each layer in the graphene and affects the band structure [17,58]. Here, we used the same method as in Refs. [17], where it was assumed that the external electric field diminishes exponentially with the screening length $\lambda$, which is a fitting parameter to be experimentally determined. We estimated it to be about 0.43 in our previous work on the Landau level structure in AB-stacked multilayer graphene in which we measured samples with a single gate electrode [17,36]. The resistance ridges observed in the present experiment were best explained for $\lambda \sim 0.45$ nm. Here, we assumed the dielectric constant in the graphene to be $\varepsilon/\varepsilon_0 = 2.0$.

E Calculation of conductivity at zero magnetic field.

The Drude conductivity was calculated by using the numerically calculated dispersion relations. The resistivity was then determined by taking the reciprocal of the conductivity. A constant relaxation time was assumed. For a small electric field $E_x$ applied in the $x$-direction, the solution of the Boltzmann equations is simply approximated at low temperature by shifting the wave number ($k$) of all the existing electrons by $-eE_x\tau$. Thus, the conductivity is proportional to the sum of the group velocities for all of the filled electronic states. To make a comparison with the experiment, we took energy broadening of the distribution function into account; this would possibly arise for various reasons, *e.g*, scattering, inhomogeneity, etc. We assumed that the derivative of the distribution function with respect to energy is simply a Gauss function with a standard deviation, $\Gamma/\sqrt{2}$. The details of the distribution function would not change the important feature of the simulation.

## Acknowledgements

This work was supported by KAHENHI No. 25107003 from MEXT Japan.

Figure captions

Fig. 1 (a) Simplified dispersion relations for AB-stacked multilayer graphene. Graphene with an odd number of layers consists of bilayer band(s) and a monolayer band. Graphene with an even number of layers consists of bilayer band(s). (b) Optical micrograph of encapsulated graphene sample with top and bottom gate electrodes (top). The bar is 10 μm. (c) Illustration of vertical structure of encapsulated graphene in the effective sample area. G means graphene. TG means the top gate electrode, and BG means the conducting Si substrate. (d) Back gate voltage ($V_b$) dependence of resistivity of AB-stacked 5-layer graphene sample for different top gate voltages ($V_t$). $V_t$ was varied between -10 and 10 V in 2 V steps.

Fig. 2 Top and bottom gate voltage dependence of resistivity in AB-stacked 5-layer graphene.
(a) Map of resistivity as a function of $V_t$ and $V_b$. $T = 4.2$ K. $B = 0$ T. (b) Map of resistivity as a function of $n_{tot}$ and $D_\perp$. (c) Similar map for $d\rho/dn_{tot}$.

Fig. 3 Band structure of multilayer graphene in perpendicular electric field.
(a) Dispersion relations in AB-stacked 5-layer graphene. $D_\perp$ was varied from 0 to $0.802 \times 10^{-7}$ and $4.81 \times 10^{-7}$ cm$^{-2}$As. (b) Similar results for AB-stacked 4-layer graphene. Bands are labeled $\alpha_e$, $\beta_e$, $\gamma_e$, $\alpha_h$, $\beta_h$, and $\gamma_h$. The characteristic points in the bands are labeled a–c and a'-b'. The right inset shows the definition of the Slonczewski-Weiss-McClure (SWMcC) parameters. The SWMcC parameters of graphite were used for the calculations ($\gamma_0$=3.19 eV, $\gamma_1$=0.39 eV, $\gamma_2$=−0.02 eV, $\gamma_3$=0.3 eV, $\gamma_4$=0.044 eV, $\gamma_5$ =0.038 eV, and $\Delta_p$=0.037 eV).

Fig 4. Resistance ridges and characteristic band points in AB-stacked 5-layer graphene.
(a) Map of $d\rho/dn_{tot}$ as a function of $n_{tot}$ and $D_\perp$. The left panel shows results from the experiment, and the right panel is the numerical calculation with $\Gamma = 3$ meV. Resistance ridges b, c and b' correspond to the positions in the dispersion relation in Fig. 3(a). The areas surrounded by the red lines indicate the measured area in the experiment. (b) Similar plot for numerical simulation with $\Gamma = 1$ meV. Resistance ridges originating from the monolayer-like band (a and a') are discernible at large values of $n_{tot}$.

Fig. 5 Intrinsic resistance ridges for AB-stacked 7-layer graphene

(a) Map of $\rho$ as a function of $V_b$ and $V_t$. $T = 4.2$ K. $B = 0$ T. (b) Replot as a function of $n_{tot}$ and $D_\perp$.

Fig. 6  Band structure of multilayer graphene in perpendicular electric field.
(a) Dispersion relations of AB-stacked 7-layer graphene. From left to right, $D_\perp$ was varied from 0 to $0.802 \times 10^{-7}$ and $4.81 \times 10^{-7}$ cm$^{-2}$As. (b) Dispersion relations of AB-stacked 6-layer graphene. Bands are labeled $\alpha_e$, $\beta_e$, $\gamma_e$, $\delta_e$, $\alpha_h$, $\beta_h$, $\gamma_h$ and $\delta_e$. The characteristic points in the bands are labeled with a –d and a'-c'.

Fig7.  Resistance ridges and characteristic band points in AB-stacked 7-layer graphene. Map of $d\rho/dn_{tot}$ as a function of $n_{tot}$ and $D_\perp$. The left panel shows experimental results, and the right panel is a calculation with energy broadening $\Gamma = 3$ meV. b, c, d, b', and c' correspond to the positions in the dispersion relation. The areas surrounded by the red lines indicate the measured area in the experiment.

Fig. 8. Evolution of resistance ridge structure in AB-stacked multilayer graphene.
Numerically calculated maps of $d\rho/dn_{tot}$ (upper panels) and $\rho$ (lower panels) are plotted against $n_{tot}$ and and $D_\perp$. From left to right, the number of layers are 4, 5, 6 and 7. **b** stands for the ridge structure due to bilayer-like bands, and **m** stands for that due to monolayer-like bands. **MDP** stands for the resistance ridge structure arising from mini-Dirac points. $\Gamma$=1 meV, and the SWMcC parameter of graphite were used for these calculations.

Fig. 9 Landau level structure in AB-stacked 5-layer graphene
(a) Map of longitudinal resistivity $\rho_{xx}$ in AB-stacked 5-layer graphene. $D_\perp = 0$ cm$^{-2}$As. $T = 4.2$ K. Numbers show filling factors for some energy gaps. (b) Map of $d\rho_{xx}/dn_{tot}$. Red bars indicate Landau levels for the monolayer-like band, which appear as a beating of the magnetoresistance oscillations. (c) Numerically calculated energy eigenvalues for AB-stacked 5-layer graphene. Red and black lines show data for K and K' points, respectively. The SWMcC parameters of graphite were used for this calculation. (d) Map of numerically calculated density of states (DOS).

Fig. 10. Landau level structure in AB-stacked 7-layer graphene
(a) Map of longitudinal resistivity of AB-stacked 7-layer graphene. $D_\perp = 0$ cm$^{-2}$As. $T = 4.2$ K. Numbers show filling factors for some energy gaps.(b) Map of $d\rho_{xx}/dn_{tot}$. (c) Numerically calculated energy eigenvalues. Red and black lines show data for K and K' points, respectively. The SWMcC parameters of graphite were used. (d) Map of numerically calculated density of states.

Fig 1

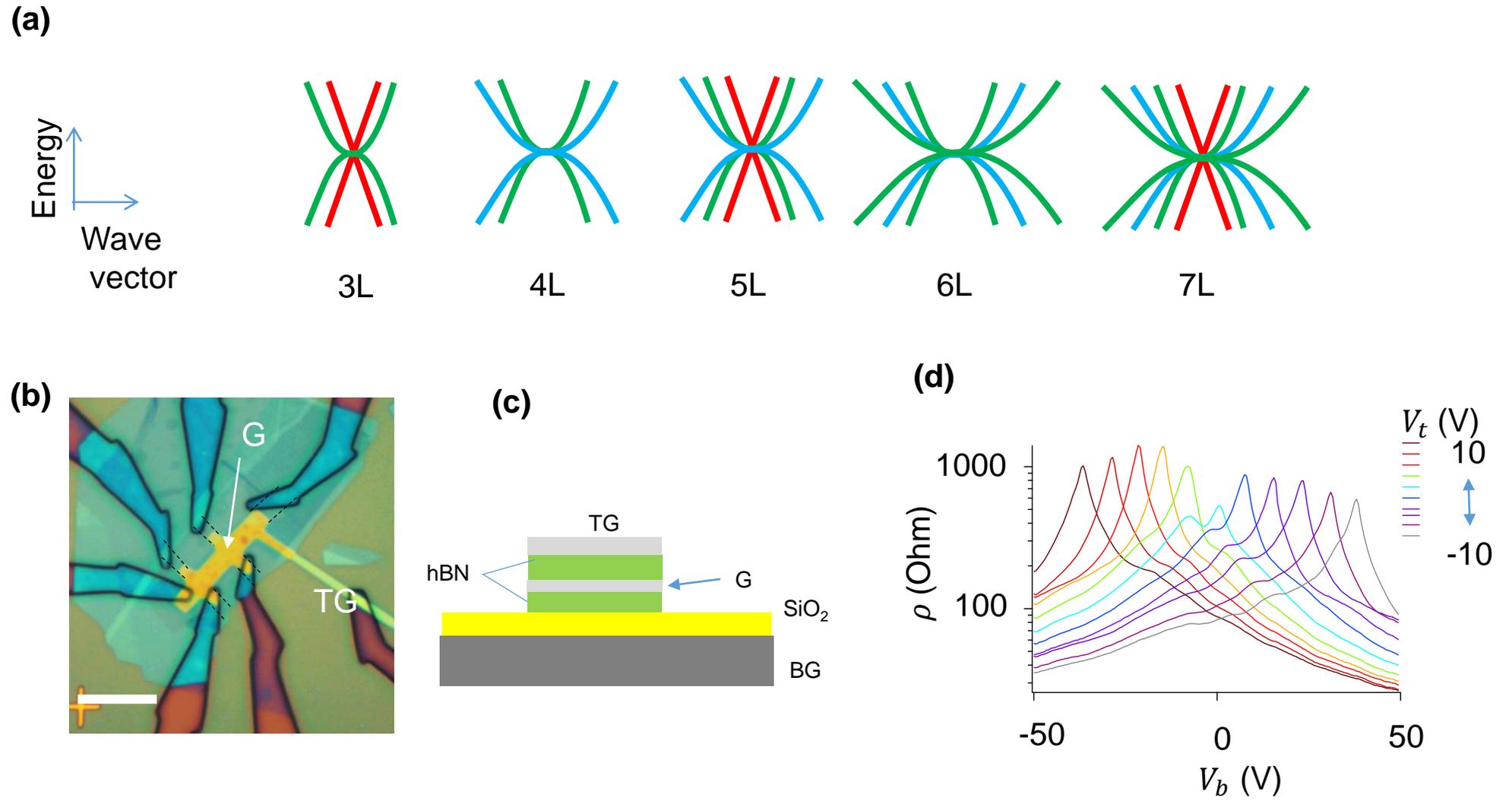

Fig 2

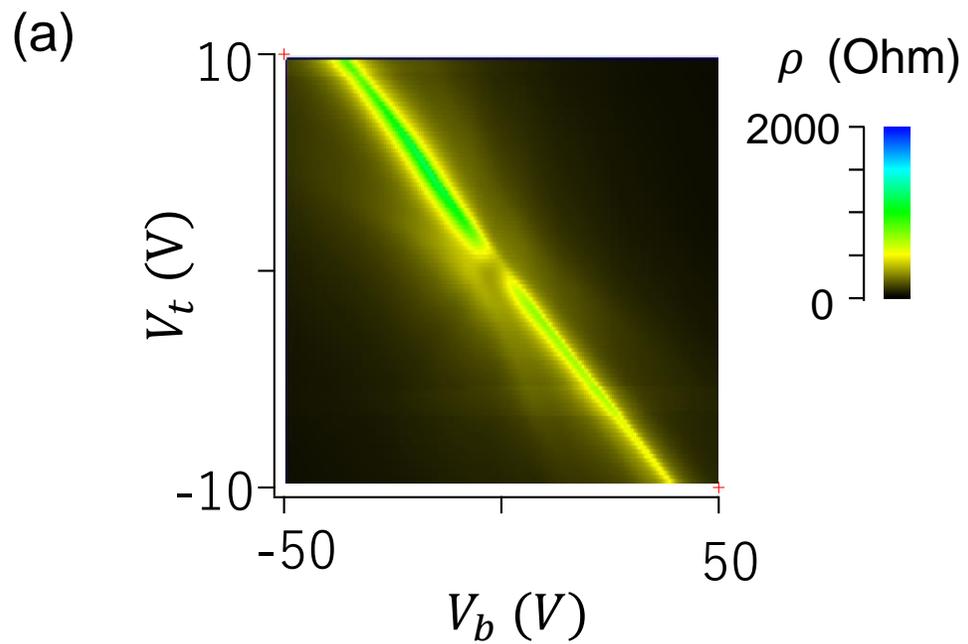
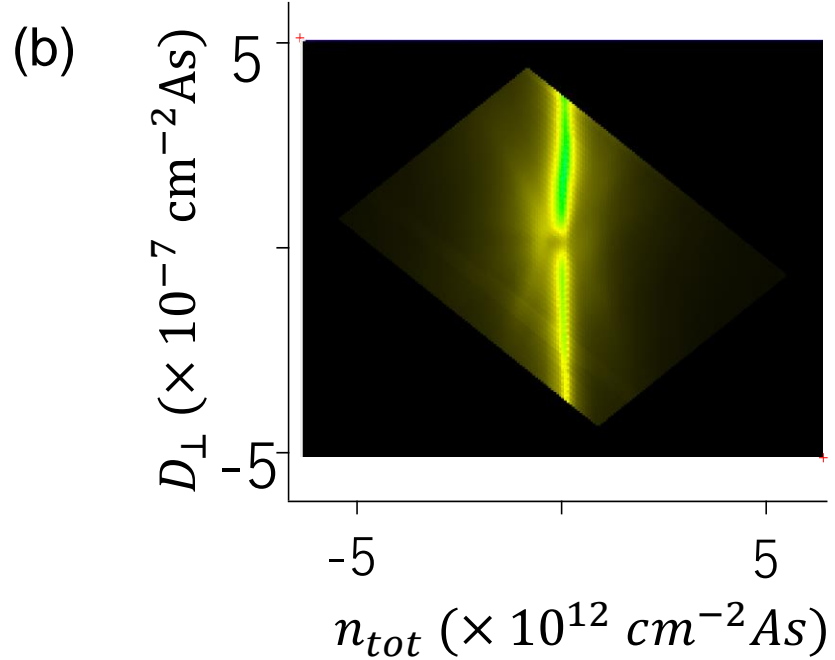
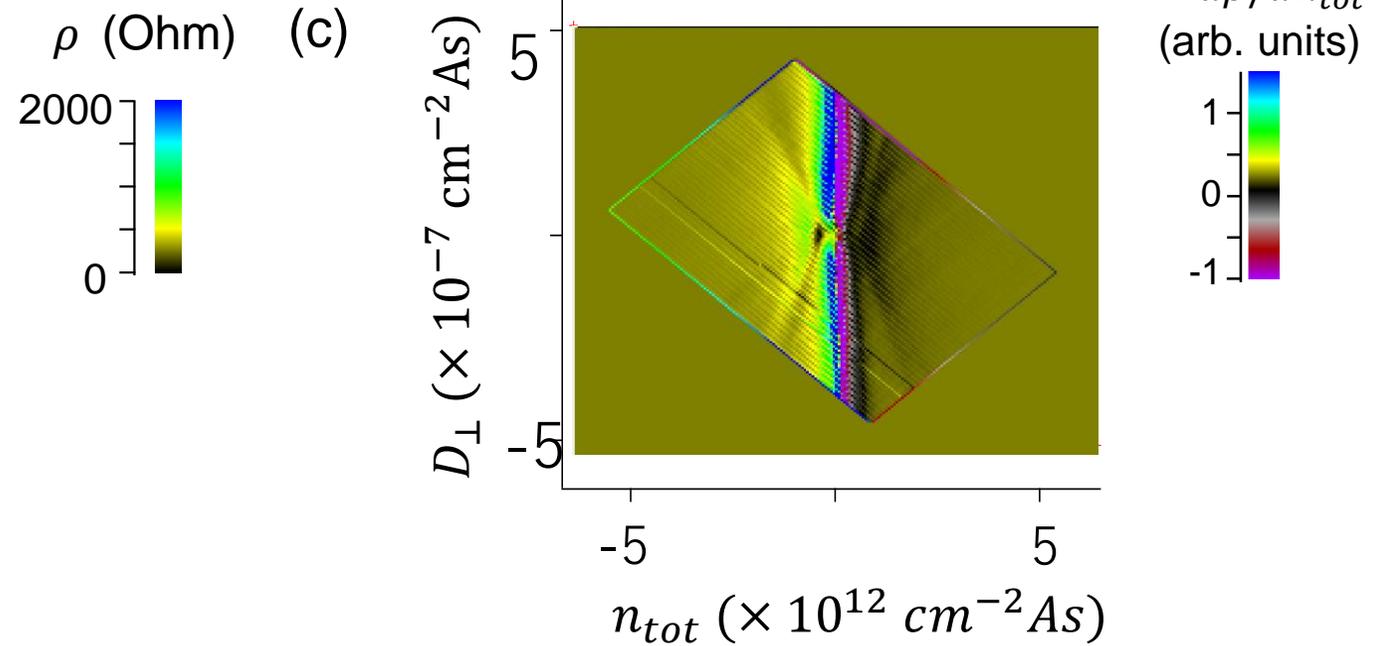

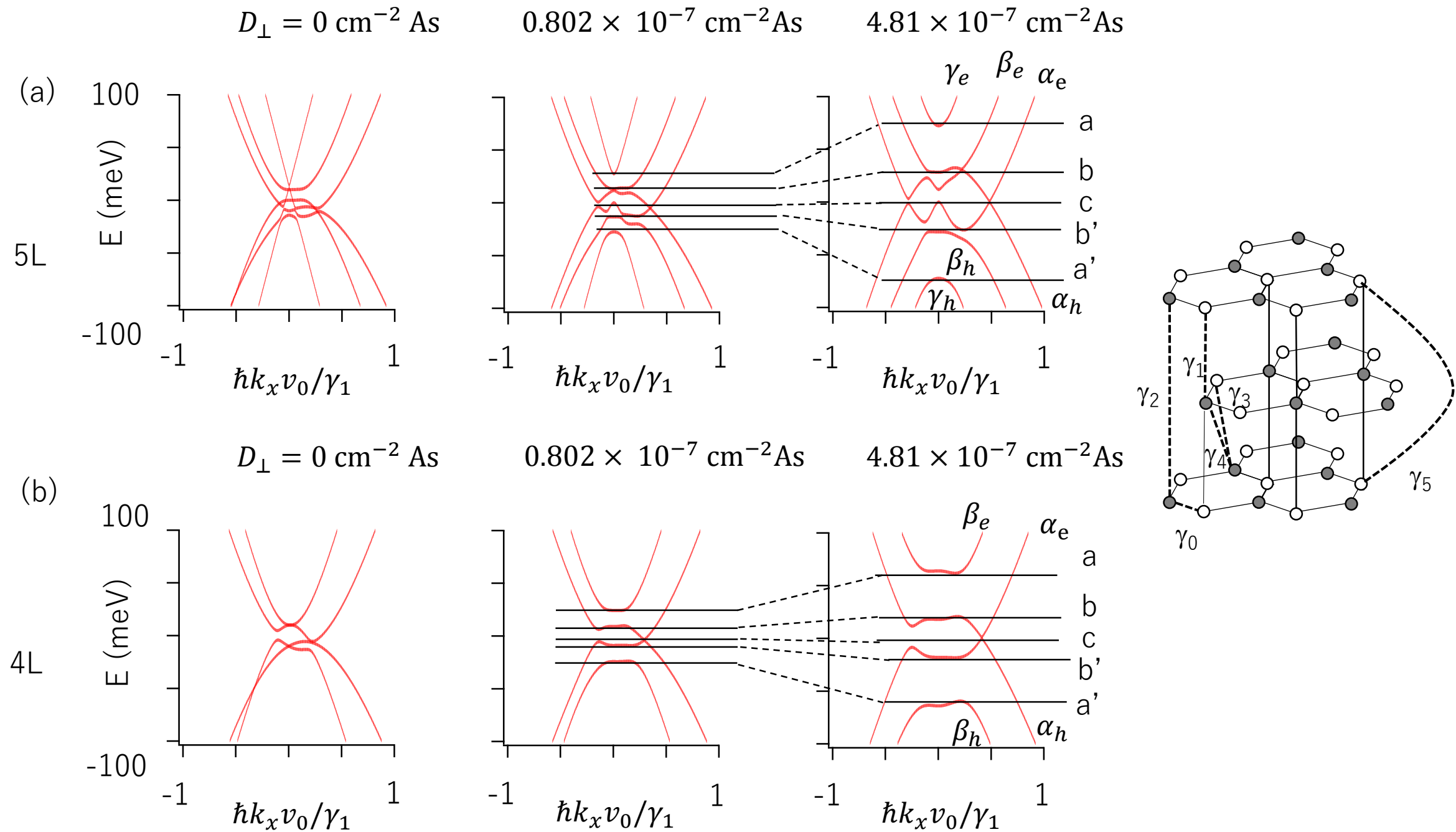

Fig 4

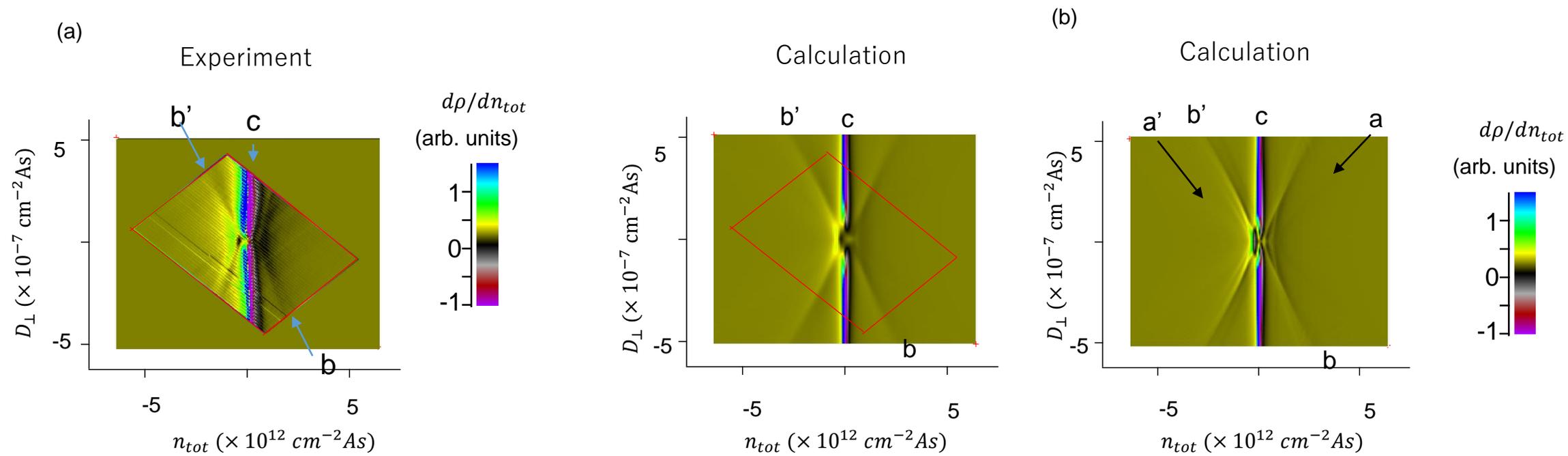

Fig5

(a) 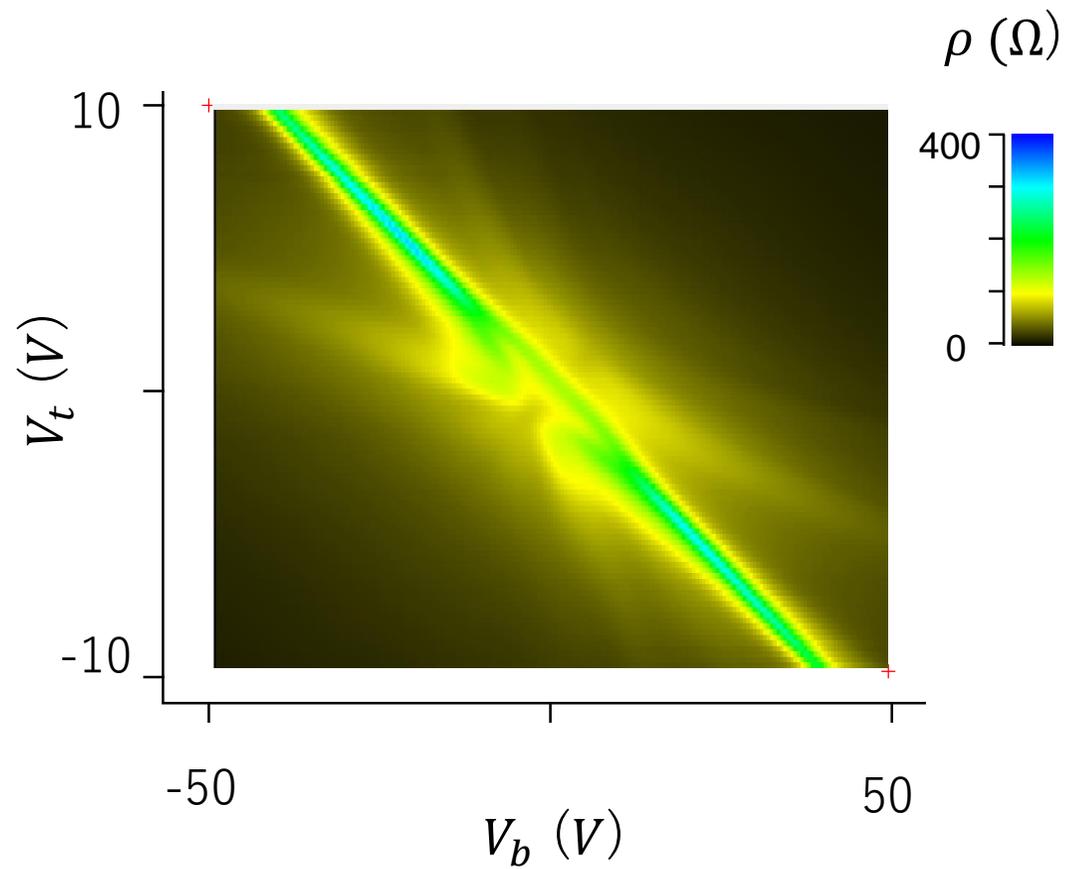

(b) 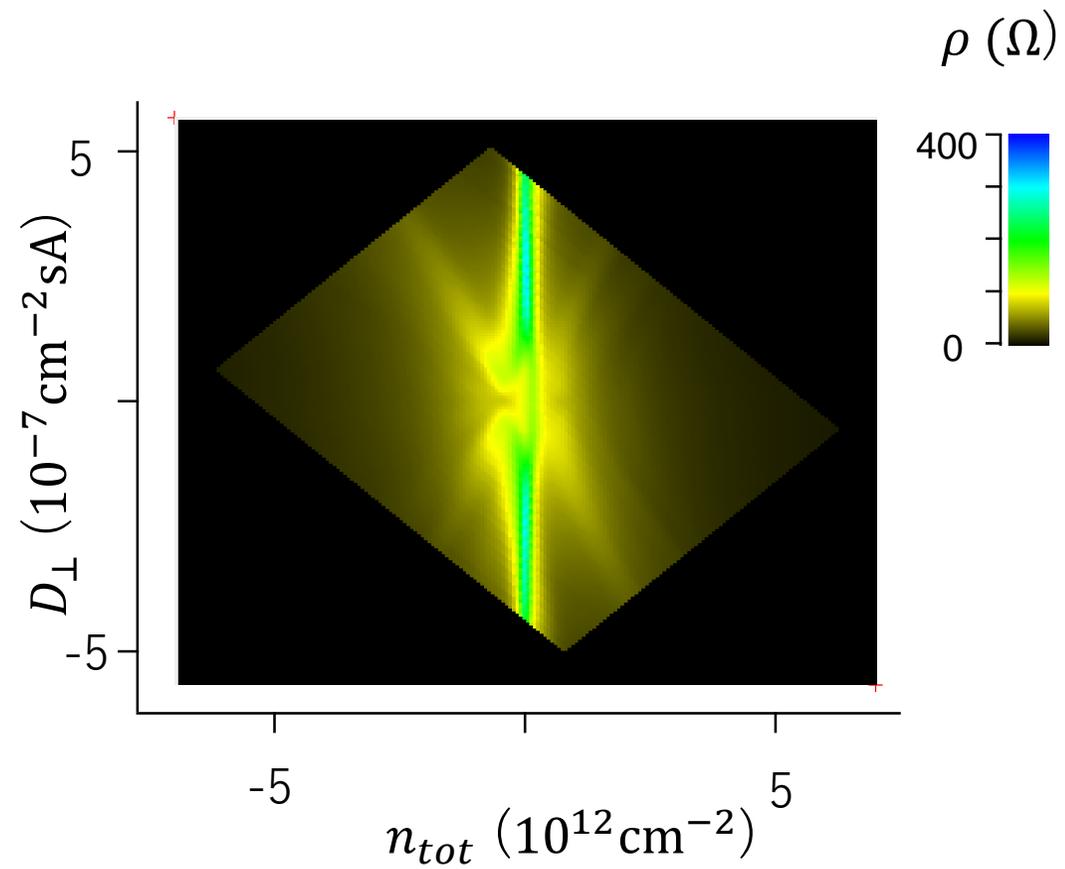

Fig 6

(a)

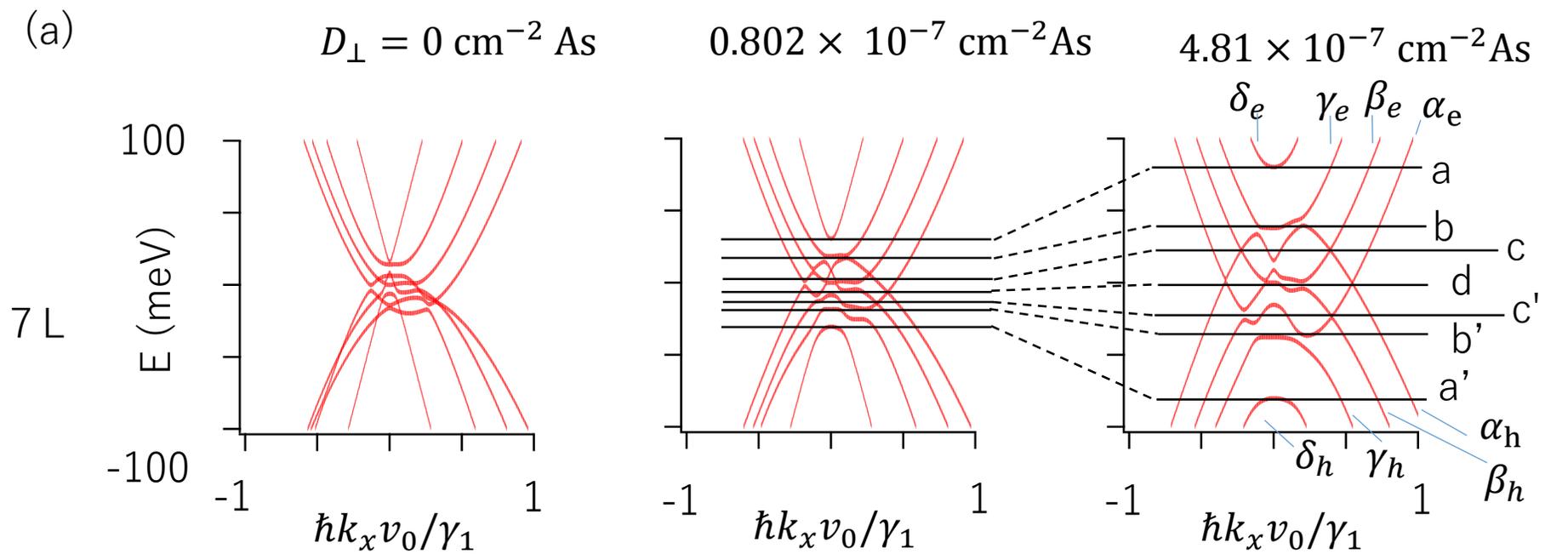

7 L

(b)

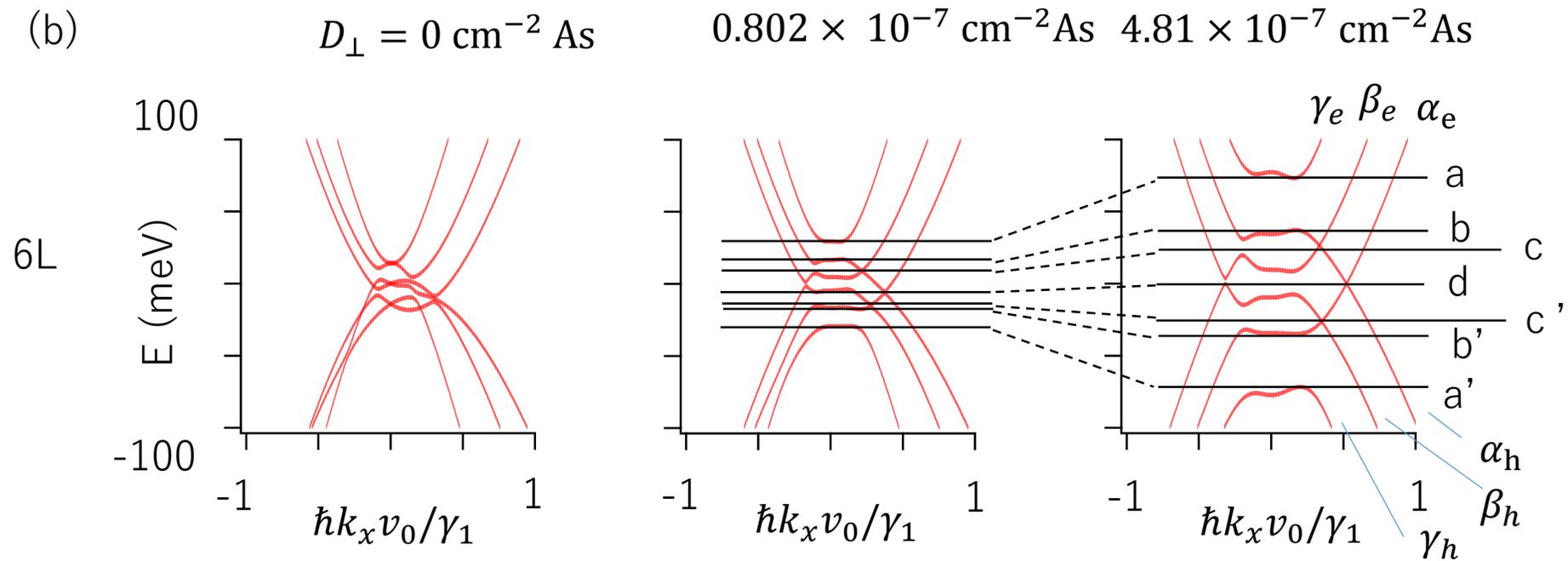

6L

Fig 7

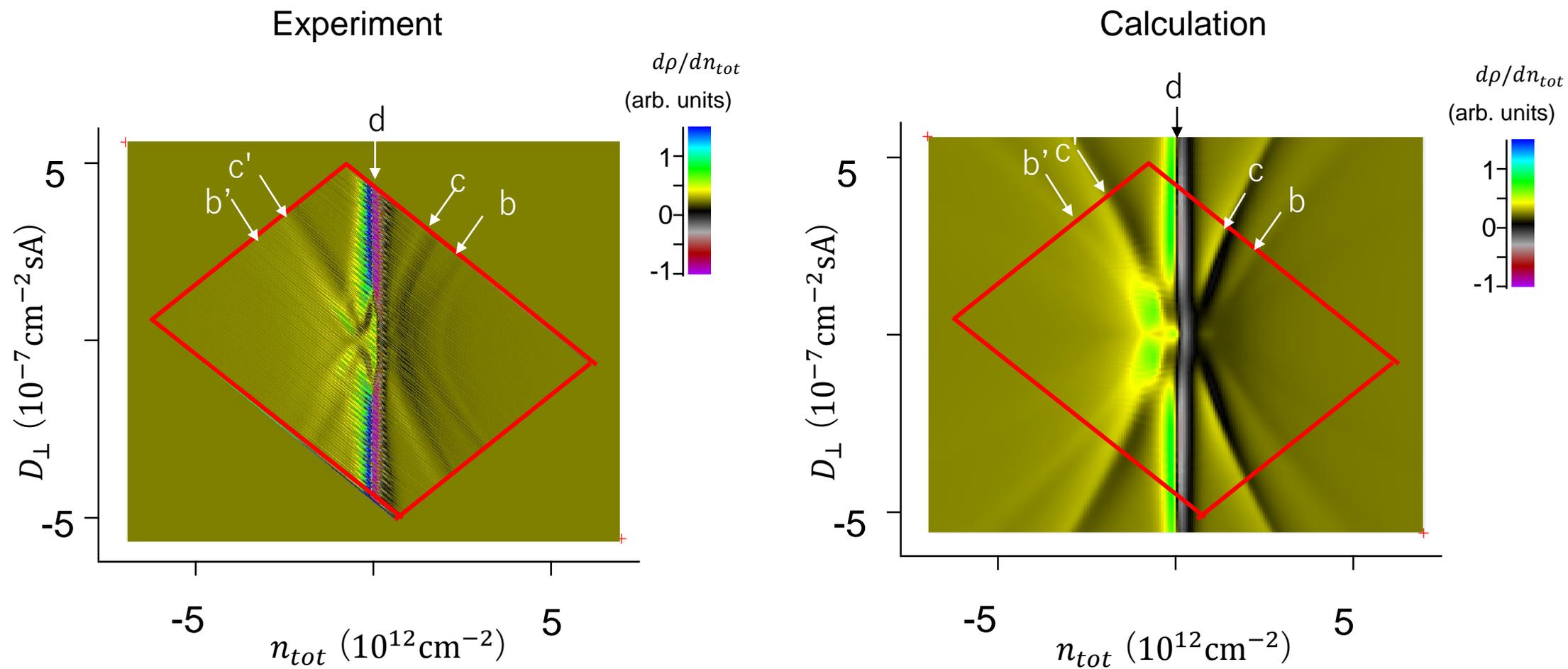

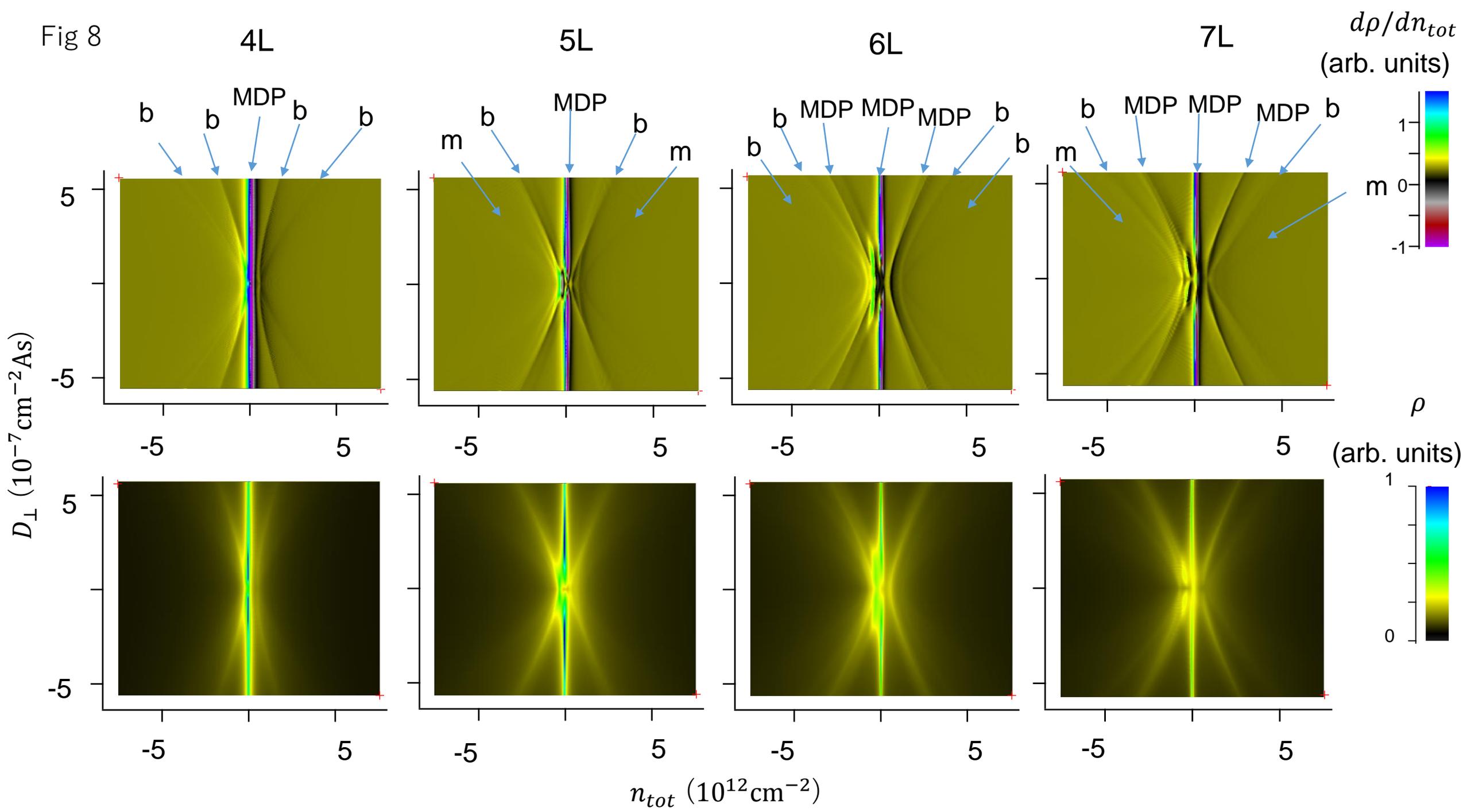

Fig 8

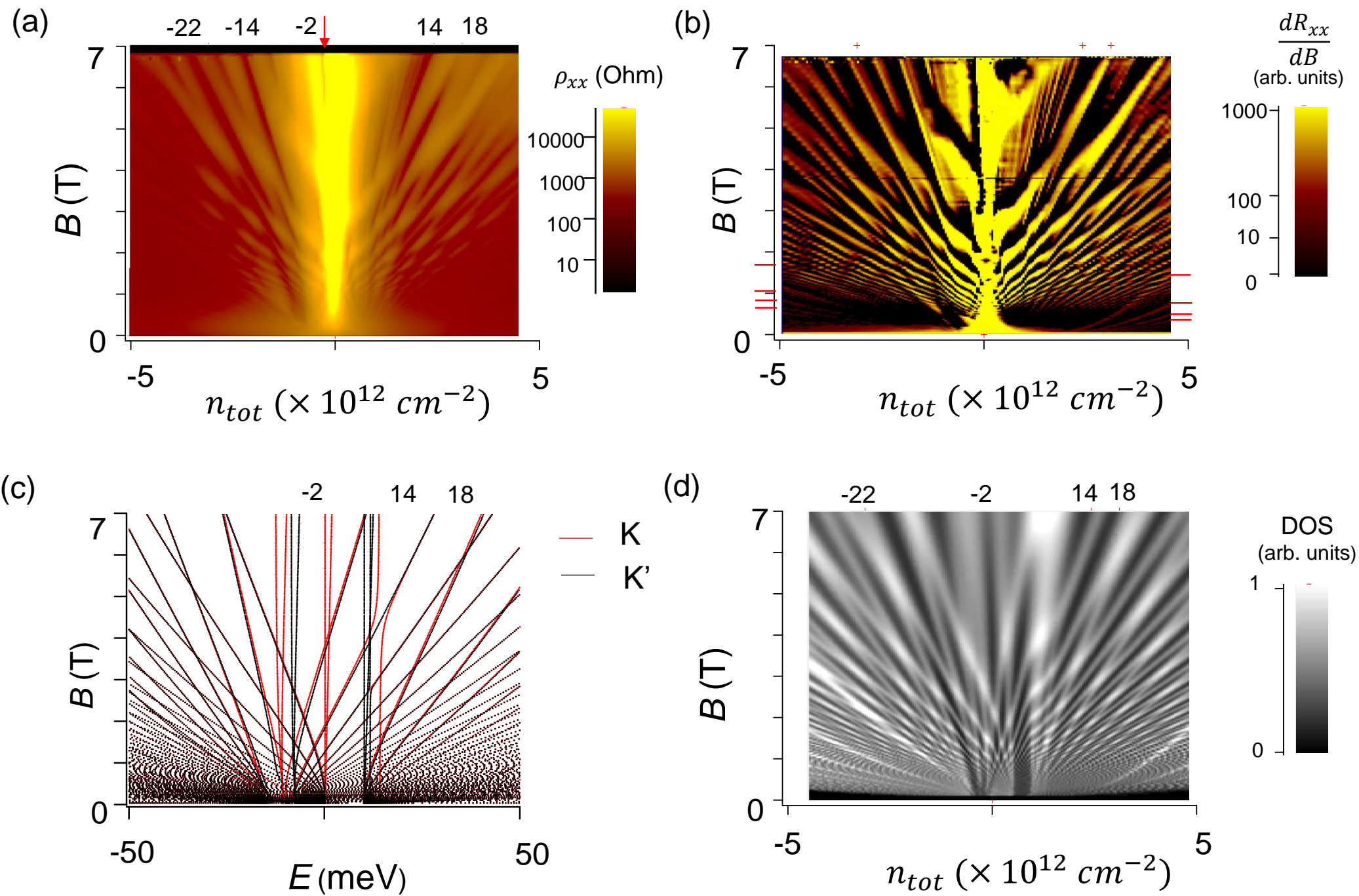

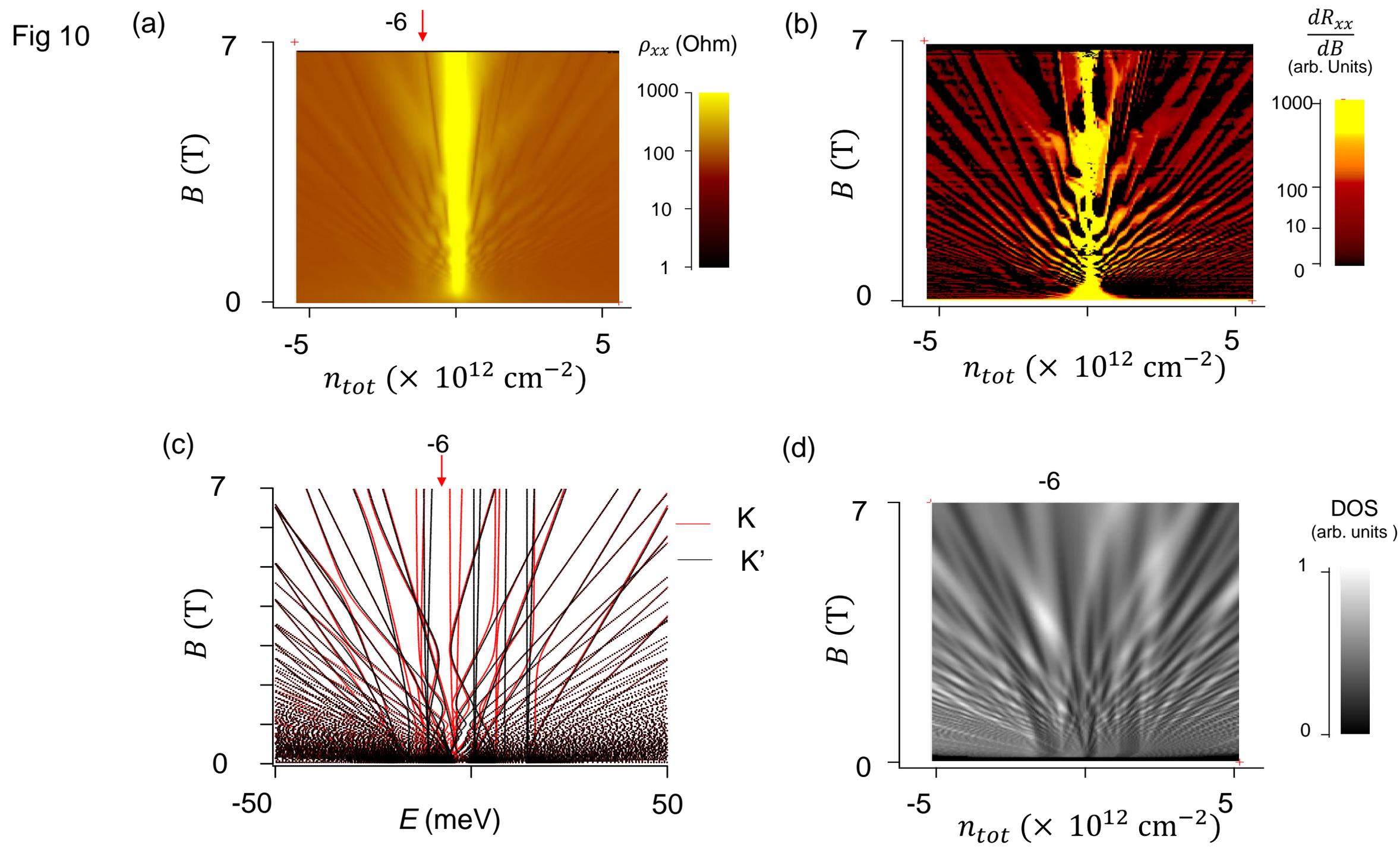

Fig 10